\documentclass[twocolumn,showpacs]{revtex4}
\begin{document}

\title{Quantum Drag Forces on a Sphere Moving Through a Rarefied Gas}

\author{D. Drosdoff}\email{drosdoff.d@neu.edu}
\affiliation{Physics Department, Northeastern University, Boston MA USA}
\author{A. Widom}
\affiliation{Physics Department, Northeastern University, Boston MA USA}
\author{Y. Srivastava}
\affiliation{Physics Department, Northeastern University, Boston MA USA}
\affiliation{Physics Department \& INFN, University of Perugia, Perugia Italy}

\begin{abstract}
As an application of quantum fluid mechanics, we consider the drag force exerted
on a sphere by an ultra-dilute gas. Quantum mechanical diffraction scattering theory
enters in that regime wherein the mean free path of a molecule in the gas is large
compared with the sphere radius. The drag force is computed in a model specified by
the ``sticking fraction'' of events in which a gaseous molecule is adsorbed by the
spherical surface. Classical inelastic scattering theory is shown to be inadequate
for physically reasonable sticking fraction values. The quantum mechanical scattering
drag force is exhibited theoretically and compared with experimental data.
\end{abstract}


\maketitle

\section{Introduction}

Quantum fluid mechanical effects\cite{LandauQFD} are very often considered to be
negligible. Except for some very special cases\cite{Kogan}, such as the study of
the superfluid phases of helium\cite{Feynman, Donnelly}, quantum fluid mechanics
is rarely considered. Here we wish to consider an important exception to this
rule; i.e. the drag force exerted on a moving sphere by a highly rarified gas.
We wish to consider the case in which the mean free path of a gas molecule is
large on the length scale of the sphere radius. For example,
a very rarefied gas\cite{Schaaf2} exists above the upper atmosphere. Meteors
or spaceships on first entering such an atmosphere\cite{Sharipov,Liffman} may
approximate the situation to be studied in this work.

A rarified gas will exert a drag force on a moving sphere. If the mean free path
of a molecule in the gas is small compared with the radius of the sphere,
then the drag is due to the viscosity of the gas. If the gas is further diluted so that
the mean free path of a molecule is much larger than the sphere radius, then
the drag force in a kinetic theory picture depends on the notion of a sticking
fraction \begin{math} f \end{math}, i.e. the fraction \begin{math} 0<f<1 \end{math}
of molecules incident on a surface which stick and thermalize to the temperature
of the sphere before later evaporating. We shall later presume that those
molecules which do not stick to the surface are specularly reflected.
The central result of such a Knudsen model\cite{Knudsen:1911} is a relationship between
the sticking fraction and the slip drag force on the sphere.

In past treatments of the scattering of molecules off the sphere, the classical
scattering theory has been employed\cite{Zhigang}. A central result of our work
is that the classical scattering theory is inadequate and should be replaced by
quantum scattering theory. The drag coefficient is proportional to the transport
cross section for molecules to scatter off the sphere. Due to diffraction effects,
the quantum mechanical cross section will differ appreciably from the classical
cross section. The incoming molecules are sufficiently large and fast for
their quantum wavelengths to be very small on the scale of the sphere radius.
This is a necessary, but not sufficient condition for the classical limit to
be obtained. The classical limit still fails to hold true due to diffractive
shadow scattering which persists to even the smallest wavelengths. Diffraction
effects increase the total cross section by roughly a factor two. The factor is
exactly two for purely specular reflection. Millikan\cite{Millikan} made a series
of experiments measuring the drag force on oil droplets by rarefied gases and
found an effective transport cross section
given by \begin{math} \sigma_m\approx 1.37\pi a^2\end{math}
wherein \begin{math} a \end{math} is the sphere radius. It is not very easy to
understand why such an experimental cross section is larger than the standard
classical geometric cross section \begin{math} \pi a^2 \end{math}.
Unfortunately we do not know of any recent experiments,
which attempt to measure the drag forces on spheres in a similar Knudsen
regime\cite{Li, Allen}. We thereby use the reliable Millikan experimental
results to compare with theory.

In Sec.\ref{Classical drag}, firstly we review the classical kinetic theory
of the drag coefficient on a sphere for the case of completely elastic
classical specular reflection. Secondly, the completely absorptive limit is
briefly discussed. In Sec.\ref{Quantum drag} we firstly review the
kinetic theory of the drag coefficient on a sphere for the case of completely
elastic {\em quantum} specular reflection. The quantum kinetic theory for
the drag coefficient will be calculated. Secondly, the completely {\em quantum}
absorptive limit is briefly discussed. In Sec.\ref{model} a general sticking
fraction model will be discussed. The results obtained thereof will be
compared to experiment. In the concluding Sec.\ref{conc}, the implication of
these results for the picture of quantum turbulent back-flow will be discussed.
Quantum diffraction effects lead to a strong forward peak in the differential
cross section. In quantum hydrodynamic terms, the forward scattering peak
translates into a thin quantum trail of fluid which will flow behind a moving
sphere. The trail is to those particles which undergo specular reflection
from the surface whose probability exhibits strong diffraction and interference
effects.

\section{Classical Drag Force on a Sphere \label{Classical drag}}

Many of the kinetic models applied to drag forces in ultra-rarefied gases involve
the notion of a sticking fraction \begin{math} f \end{math} defined as
follows: (i) \begin{math} (1-f) \end{math} is the probability that a molecule
will elastically scatter off the surface in a specular fashion.
(ii)  \begin{math} f \end{math} is the probability that a molecule will stick
to the surface and thermalize to the substrate temperature before finally being
evaporated back into the gas. Below, the drag force on a sphere of radius $a$ will be
reviewed using purely classical scattering theory and classical kinetic theory.
Two opposite limits will be discussed; namely (A) the purely elastic specular
reflection limit, i.e. \begin{math} f=0 \end{math}, and (B) the purely inelastic
absorption limit, i.e. \begin{math} f=1 \end{math}.

\subsection{Classical Elastic Specular Reflections}

Consider a particle moving in the gas in the direction of a unit vector
\begin{math} {\bf u} \end{math}. If the particle hits the sphere and specularly
scatters through an angle \begin{math} \theta  \end{math}, then the component
of the momentum transfer along the original direction is given by
\begin{equation}
{\bf u}\cdot ({\bf p}_i-{\bf p}_f)=|{\bf p}|(1-\cos \theta ).
\label{cdrag0}
\end{equation}
If one now has \begin{math} n \end{math} gas particles of mass \begin{math} m \end{math}
per unit volume with a distribution of momenta \begin{math} {\bf p}=p{\bf u} \end{math},
then the momentum transfers give rise to a drag force
\begin{equation}
{\bf F} = \left< n{\bf p}{p\over m}\int (1-cos\theta )d\sigma \right>_{\bf p}.
\label{cdrag1}
\end{equation}
In Eq.(\ref{cdrag1}), \begin{math} d\sigma \end{math} is the differential cross
section of scattering from the sphere into a solid angle
\begin{math} d\Omega \end{math} about the angle \begin{math} \theta  \end{math}.
The brackets \begin{math}\left<\ldots  \right>_{\bf p} \end{math} denote
averaging over all momenta ${\bf p}$ with a Maxwellian distribution
corresponding to a mean gas ``wind velocity'' \begin{math} {\bf v} \end{math}.
The transport cross section is usually defined\cite{Ferziger} as
\begin{equation}
\sigma_1=\int (1-cos\theta )d\sigma,
\label{cdrag2}
\end{equation}
so that drag force may be written
\begin{equation}
{\bf F} = \left< n{\bf p}{p\over m}\sigma_1\right>_{\bf p}.
\label{cdrag3}
\end{equation}

For classical specular reflection from a spherical surface,
\begin{equation}
\left[\frac{d\sigma }{d\Omega}\right]_{specular}=\frac{a^2}{4}
\ \ ({\rm classical\ hard\ sphere}),
\label{cdrag2a}
\end{equation}
the transport cross section is the
same as total cross section(\cite{Landau}); i.e.
\begin{equation}
\sigma_1=\sigma=\pi a^2\ \ ({\rm classical\ elastic\ hard\ sphere}).
\label{cdrag4}
\end{equation}
The drag force for the above cross sections
is\cite{Shidlovskiy,Schaaf,Cercignani,Mayday,Bailey,Brown}
\begin{eqnarray}
{\bf F}&=& \frac{\pi a^2n}{m}\int p{\bf p}
e^{-({\bf p}-m{\bf v})^2/ 2mk_BT}
\left[\frac{d^3 {\bf p}}{(2\pi mk_BT)^{3/2}}\right]
\nonumber \\
F&=&{1\over 2}\rho \pi a^2C_Fv^2\ \ \ (\rho=mn),
\label{cdrag5}
\end{eqnarray}
where
\begin{equation}
C_F = {e^{-s^2}\over \sqrt \pi s^3}(2s^2+1) +
{{\rm erf}(s)\over 2s^4}(4s^4+4s^2-1),
\label{cdrag6}
\end{equation}
\begin{equation}
s=v\sqrt{\frac{m}{2k_BT}}\ ,
\label{cdrag7}
\end{equation}
and \begin{math} {\rm erf}(s)  \end{math} is the error
function\cite{AandS}. When the velocity of the object is low,
then the drag force obeys\cite{Epstein}
\begin{equation}
\lim_{v\to 0}\left[\frac{F}{v}\right]
=\frac{4}{3}(\pi a^2)\rho\overline{c},
\ \ {\rm where}\ \ \bar{c}=\sqrt{{8k_BT \over \pi m}}\ .
\label{cdrag8}
\end{equation}
The mean speed of a gas particle is denoted by \begin{math} \bar{c} \end{math}.

\subsection{Classical Purely Inelastic Absorption}

By purely inelastic classical absorption we mean that
any incoming particle whose impact parameter is less than
the sphere radius sticks to the spherical surface
with probability one; i.e. \begin{math} f=1 \end{math}.
The particle may much later be re-emitted
after thermal equilibrium with the sphere is established.
This kind of evaporation implies the re-emission of gas particles
with a Maxwellian distribution with zero mean velocity
in the reference frame of the sphere. The spherical
nature of the re-emission implies the equality of the
transport cross section in Eq.(\ref{cdrag3}) and total cross section.
Specifically,
\begin{equation}
\sigma_1(f=1)=\sigma_{in}=\pi a^2\ \ \ {\rm classical}.
\label{cdrag9}
\end{equation}
It then follows that the drag force on a sphere due to purely
classical elastic collisions coincides with the drag force
due to purely classical inelastic collisions\cite{Epstein}.
The drag force is again,
\begin{equation}
F={1\over 2}\rho \pi a^2C_Fv^2,
\label{cdrag10}
\end{equation}
wherein \begin{math} C_F \end{math} is defined in Eq.(\ref{cdrag6}).

\section{Quantum Drag Force on a Sphere \label{Quantum drag}}

The elastic amplitude for a gas molecule to scatter off a sphere may
be expanded into partial waves
\begin{equation}
f(\theta ,p)=\left(\frac{\hbar }{2ip}\right)\sum_{l=0}^\infty
(2l+1)[S_l(p)-1]P_l(\cos\theta).
\label{QD1}
\end{equation}
The total cross section can be decomposed into an elastic plus
an inelastic part
\begin{equation}
\sigma_{tot}(p)=\sigma_{el}(p)+\sigma_{in}(p).
\label{QD2}
\end{equation}
The total cross section follows from the ``optical theorem''
\begin{eqnarray}
\sigma_{tot}(p)&=&\left(\frac{4\pi \hbar }{p}\right){\Im }mf(0,p)
\nonumber \\
\sigma_{tot}(p)&=&
\left(\frac{2\pi \hbar^2 }{p^2}\right)
\sum_{l=0}^\infty (2l+1)[1-{\Re}eS_l(p)].
\label{QD3}
\end{eqnarray}
The elastic cross section is determined by
\begin{eqnarray}
\frac{d\sigma_{el}}{d\Omega}&=&|f(\theta ,p)|^2
\nonumber \\
\sigma_{el}(p)&=&\int|f(\theta ,p)|^2{d\Omega}
\nonumber \\
&=&\left(\frac{\pi \hbar^2 }{p^2}\right)
\sum_{l=0}^\infty (2l+1)|1-S_l(p)|^2,
\label{QD4}
\end{eqnarray}
so that
\begin{equation}
\sigma_{in}(p)=\left(\frac{\pi \hbar^2 }{p^2}\right)
\sum_{l=0}^\infty (2l+1)[1-|S_l(p)|^2].
\label{QD5}
\end{equation}
Thus, the probability \begin{math} w^{el}_l \end{math} of elastic
scattering and the probability \begin{math} w^{in}_l \end{math}
of inelastic scattering in a given partial wave are determined,
respectively, by
\begin{equation}
w^{el}_l=|S_l(p)|^2\ \ {\rm and}\ \ w^{in}_l=1-|S_l(p)|^2.
\label{QD6}
\end{equation}

\subsection{Quantum Pure Elastic Scattering}
If the probability of elastic scattering in a partial wave is unity,
then one may define phase shifts \begin{math} \{\delta_l(p)\} \end{math}
via
\begin{equation}
w^{el}_l=1\ \ {\rm implies}\ \ S_l(p)=e^{2i\delta_l(p)}.
\label{QD7}
\end{equation}
The elastic cross section Eq.(\ref{QD4}) yields
the quantum result
\begin{equation}
\sigma(p)=\left(\frac{4\pi \hbar^2}{p^2}\right)
\sum_{l=0}^\infty (2l+1)\sin^2\delta_l(p).
\label{QD8}
\end{equation}
The quantum mechanical effect of the drag force is determined by the
transport scattering cross section \begin{math} \sigma_1(p) \end{math}.
Eqs.(\ref{cdrag3}), (\ref{QD4}) and (\ref{QD7}) imply\cite{Ziman}
\begin{equation}
\sigma_1(p) = {(4\pi \hbar^2)\over p^2} \sum_{l=0}^\infty l
\sin^2\left[\delta_{l-1}(p) -\delta_l(p)\right].
\label{qdrag1}
\end{equation}
The hard sphere phase shifts\cite{Newton} are
\begin{equation}
\tan\delta_l(p) = {{j_l(pa/\hbar)}\over{n_l(pa/\hbar)}};
\label{qdrag2}
\end{equation}
$j_l$ and $n_l$ are, respectively, the spherical Bessel and
Neumann functions.

In the high energy limit \begin{math} pa>>\hbar  \end{math};
Firstly, the asymptotic form of the phase shift is given by
\begin{equation}
\delta_l(p) \to (-pa/\hbar) + (l\pi /2) \ \
{\rm as}\ \ p\to \infty.
\label{qdrag3}
\end{equation}
Secondly, the partial wave summation cuts off at $\hbar l\approx pa$,
where $a$ is the radius of the sphere and $p$ is the particle momentum.
By inserting Eq.(\ref{qdrag3}) into Eq.(\ref{qdrag1}) one obtains
\begin{equation}
\lim_{p\to \infty }\sigma_1(p) = 2\pi a^2.
\label{qdrag4}
\end{equation}
Comparing the classical specular reflection transport coefficient
in Eq.(\ref{cdrag4}) with the quantum specular reflection Eq.(\ref{qdrag4})
one finds the drag force ratio
\begin{eqnarray}
\frac{\sigma_1({\rm quantum\ specular})}{\sigma_1({\rm classical\ specular})}
&=&2,
\nonumber \\
\frac{F({\rm quantum\ specular})}{F({\rm classical\ specular})}
&=&2.
\label{qdrag5}
\end{eqnarray}

The argument for the famous factor of two\cite{Sakurai} between the
quantum and classical cross sections is that there exists an
interference of amplitudes between the scattered wave and the
incoming wave. This interference creates a peak in the
forward direction.  This effect is closely analogous to Fresnel
diffraction in optics, wherein the limit to geometric optics
cannot really be achieved. Our point here is that the so called
classical limit \begin{math} \hbar \to 0 \end{math} cannot really
be achieved because of diffraction effects when the particle
scatters off the sphere. From Eq.(\ref{cdrag3}) one gets the
final Eq.(\ref{qdrag5}). The drag force on a sphere due to elastic
scattering is twice as large as the classical value if quantum mechanical
diffraction effects are taken into account.

\subsection{Quantum Pure Absorptive Scattering}

Pure absorptive scattering takes place when the inelastic cross section is at
maximum.  Eqs.(\ref{QD5}) and (\ref{QD6}) imply that pure absorptive scattering in
the $l^{th}$ partial wave occurs\cite{Lifshitz} when  \begin{math} S_l=0 \end{math}.
If all the partial waves scatter in a purely absorptive manner, then
\begin{math} S_{(l < ka)}\approx 0 \end{math} and
\begin{math} S_{(l > ka)}\approx 1 \end{math}. The total cross section becomes
\begin{math} \sigma_{tot}\approx 2\pi a^2 \end{math}.
The elastic cross section is thereby equal to the inelastic cross
section; i.e.
\begin{equation}
\sigma_{el}=\sigma_{in}=\pi a^2\ \ {\rm and}\ \ \sigma_{tot}=2\pi a^2.
\label{qdrag6}
\end{equation}
As in the classical result, the inelastic cross section is equal to
the inelastic transport cross section.  The ratio between the force
due to quantum scattering and classical scattering is
\begin{eqnarray}
\frac{\sigma_1({\rm quantum\ absorptive})}{\sigma_1({\rm classical\ absorptive})}
&=&2,
\nonumber \\
\frac{F({\rm quantum\ absorptive})}{F({\rm classical\ absorptive})}
&=&2.
\label{qdrag7}
\end{eqnarray}
The drag force due to quantum purely inelastic scattering
is twice as large as the classical drag force.

\section{Sticking Fraction Model \label{model}}

For purely elastic scattering, the {\it S}-matrix eigenvalues
obey the unitary condition \begin{math} |S_l|^2=1  \end{math}.
In terms of the Heitler {\it K}-matrix eigenvalues\cite{Heitler,Newton2},
\begin{equation}
S_l=\left[\frac{1-i \pi K_l}{1+i\pi K_l}\right],
\label{sf1}
\end{equation}
the unitary condition is enforced by requiring that
\begin{math} K_l \end{math} be real.
In the most simple sticking fraction model, the inelastic processes
are described with imaginary {\it K}-matrix eigenvalues,
\begin{equation}
\pi K_l=-i\eta_l\ \ {\rm where}\ \ \eta_l\ge 0.
\label{sf2}
\end{equation}
In the \begin{math}l^{th} \end{math} partial wave, a fraction
\begin{math} w^{in}_l \end{math} of incident particles ``stick''
to the sphere and a fraction \begin{math} w^{el}_l \end{math}
specularly reflect from the sphere. In detail, Eqs.(\ref{QD6}),
(\ref{sf1}) and (\ref{sf2}) imply
\begin{eqnarray}
w^{el}_l&=&\frac{(1-\eta_l)^2}{(1+\eta_l)^2}
\ \ \ \ {\rm (non-sticking)},
\nonumber \\
w^{in}_l&=&\frac{4\eta_l}{(1+\eta_l)^2}
\ \ \ \ \ {\rm (sticking)}.
\label{sf3}
\end{eqnarray}
Finally, we presume a single value for the sticking fraction
\begin{eqnarray}
w^{in}_l&\approx &f
\ \ \ {\rm if}\ \ \ \hbar l<pa,
\nonumber \\
w^{in}_l&\approx & 0
\ \ \ {\rm if}\ \ \ \hbar l>pa.
\label{sf4}
\end{eqnarray}
Eqs.(\ref{QD4})-(\ref{QD6}) and (\ref{sf1})-(\ref{sf4}) imply the
central results of the simple sticking fraction model; i.e.
\begin{eqnarray}
\sigma^{el}&=&(\pi a^2)\left[1-\sqrt{1-f}\right]^2,
\nonumber \\
\sigma^{in}&=&(\pi a^2)f.
\label{sf5}
\end{eqnarray}
The physical significance of the central quantum scattering
Eq.(\ref{sf5}) is as follows:
(i) The cross section for the incident particle to stick to the sphere is
simply the classical cross section times sticking probability. i.e.
\begin{math} \sigma^{in}=(\pi a^2)f  \end{math} which would also be valid
in a classical scattering context. (ii) The elastic cross section
\begin{math} \sigma^{el}=(\pi a^2)\left[1-\sqrt{1-f}\right]^2\end{math}
is in part due to diffractive scattering which is a purely quantum mechanical
process having {\em no classical counterpart}. (iii) The total cross
section in the sticking fraction model is
\begin{equation}
\sigma ^{tot}=2\pi a^2\left[1-\sqrt{1-f}\right]
\ \ \ {\rm (Quantum\ Sticking)}.
\label{sf6}
\end{equation}
On the other hand, for a classical sticking model, the total cross section
is geometrical;
\begin{equation}
\sigma^{tot}_{classical}=\pi a^2
\ \ \ \ {\rm (Classical\ Sticking)}.
\label{sf7}
\end{equation}
The physical kinetics of sticking and specular reflection are such that the mean
momentum transferred to the sphere are equal for elastic and inelastic events
The friction drag coefficient is thereby
\begin{equation}
\lim_{v\to 0}\left[\frac{F}{v}\right]\equiv \Gamma
=\left(\sqrt{{128k_BT \over 9\pi m}}\right)\rho \sigma ^{tot}.
\label{sf8}
\end{equation}
The quantum expression for the drag coefficient in the sticking model
is thereby
\begin{equation}
\Gamma=(2\pi a^2)\sqrt{{128k_BT \over 9\pi m}}\left[1-\sqrt{1-f}\right].
\label{sf9}
\end{equation}

The sticking fraction for very dilute air molecules bouncing off an oil surface
had been measured by two different methods: (i) In the rolling cylinder
method\cite{VanDyke}, the drag force on one rotating cylinder due to the dilute gas
between it and another nearby concentric stationary cylinder is employed to measure
the sticking fraction \begin{math} f \end{math}. (ii) In the falling drop
method\cite{Millikan}, the drag coefficient \begin{math} \Gamma \end{math} on
a falling spherical oil droplet is measured. \begin{math} \Gamma \end{math}
then determines the sticking fraction \begin{math} f \end{math}.
The droplet should have a small radius compared with the mean free path length
of an atom in the gas. Agreement between these two methods
is obtained if the fully quantum mechanical scattering theoretical Eq.(\ref{sf9})
is used in the analysis of the data. The experimental results are shown in
the table below:
\begin{center}
\begin{tabular}{|l|c|}
\hline
Experimental Method & Sticking Fraction $f$ \\
\hline\hline
Rotating Cylinder & $0.893$ \\
Falling Droplet & $0.901$  \\
\hline
\end{tabular}
\end{center}
The close agreement corresponds to a measured total cross section
for a droplet in which the Knudsen number \begin{math} Kn\gg 1  \end{math}.
Recall the conventional definition\cite{Knudsen:1911} of this number,
\begin{equation}
Kn=\frac{\Lambda}{a}\ ,
\label{sf9a}
\end{equation}
where \begin{math} \Lambda \end{math} is the mean free path of a molecule
in the rarified gas. The experimental number deduced from Eq.(\ref{sf8}) is
\begin{equation}
\sigma ^{tot}\approx 1.37 \pi a^2
\label{sf10}
\end{equation}
from which \begin{math} f \end{math} may be deduced from Eq.(\ref{sf6}).
The experimental cross section \begin{math} \sigma ^{tot} \end{math}
is larger than the classical total cross section
\begin{math} \sigma^{tot}_{classical}=\pi a^2  \end{math}
due to quantum diffraction effects.
The agreement between the experimental methods is not {\em entirely}
satisfactory since only for one oil droplet is data available to us
in which \begin{math} Kn\approx 10^2\gg 1  \end{math}.
However, the data that does exist seems to demand quantum
diffraction effects for the scattering of a molecule off the sphere.
Further experiments would be of interest for probing the accuracy of
these results.

\section{Conclusion \label{conc}}

The drag force on a sphere moving through a highly rarefied gas with
large Knudsen number has been discussed. It was found that quantum mechanics
substantially alters the drag force. Quantum mechanics enters into the
computation via diffraction effects in the cross section for molecules
scattering from the sphere. In the extreme case, it was found that if the
molecules of the gas are completely elastic or completely
inelastic, then the cross section is twice that found if only classical mechanics
is taken into account. On the other hand, when there are both elastic and
inelastic processes then the total cross section may be only somewhat higher
than the classical result. When collisions are mostly inelastic and absorptive,
the cross sections are mainly determined by the sticking fraction
\begin{math} f \end{math} for the molecule to thermalize on the oil drop surface.
Whereas classical elastic scattering is roughly isotropic,
leaving an ``empty shadow'' behind the sphere, the elastic quantum cross section
exhibits diffraction patterns strongly peaking in the forward direction of the
gas flow. A quantum wake with a narrow stream of particles will thus appear
behind the sphere.

By independent measurements of the sticking fraction and the drag force,
the quantum mechanical theory has been experimentally shown to be a more accurate
description of the flow than the classical cross section. However, experimental
data available to us in the \begin{math} Kn\gg 1 \end{math} regime is somewhat
limited. Further experiments would be of great importance.

\end{document}